\documentclass[showpacs,amsmath,amssymb,aps,prc,floatfix,twocolumn,showkeys,nofootinbib]{revtex4-1}
\pdfoutput=1 

\usepackage{graphicx}
\usepackage{dcolumn}
\usepackage{bm}
\usepackage[mathlines]{lineno}

\usepackage{slashed}
\usepackage{hyperref}
\usepackage{url}
\usepackage{color}
\begin{document}

\title{The Constituent Counting Rule and Omega Photoproduction}
\author{Trevor Reed,  Christopher Leon, Frank Vera, Lei Guo, and Brian Raue}
\affiliation{ Department of Physics, Florida International University, Miami, Florida 33199, USA}
\date{\today}


\begin{abstract}
The constituent counting ruling (CCR) has been found to hold for numerous hard, exclusive processes. It predicts the differential cross 
section at high energies and fixed $\cos \theta_{c.m.}$ should follow $ \frac{d \sigma}{dt} \sim \frac{1}{s^{n-2}}$, where $n$ is the minimal 
number of constituents involved in the reaction. Here we provide an in-depth analysis of the reaction $\gamma p \rightarrow \omega  p$ at 
$\theta_{c.m.}\sim 90^\circ$ using CLAS data with an energy range of $s = 5 - 8$ GeV$^2$, where the CCR has been shown to work in 
other reactions. We argue for a stringent method to select data to test the CCR and utilize a Taylor-series expansion to take advantage of data 
from nearby angle bins in our analysis.  Na\"{i}vely, this reaction would have $n=9$ (or $n=10$ if the photon is in  a $q\bar{q}$ state) and we 
would expect a scaling of $\sim s^{-7}$ ($s^{-8}$). Instead, a scaling of $s^{-(9.08 \pm 0.11)}$ was observed. A careful analysis of conservation of angular momentum is proposed to explain the discrepancy, supporting the validity of CCR when applied properly.
\end{abstract}
\maketitle

\section{Introduction}
The transition from hadronic to partonic degrees of freedom is an interesting area of nuclear physics that is still not well understood. Knowing which kinematic regions have what effective degrees of freedom and how the transition between the two occurs can tell us much about quantum chromodynamics (QCD). Currently these problems are very difficult to solve purely through theoretical tools and thus experiment can be used to provide guidance.

In the early days of QCD it was recognized that one of the consequences of having partons as the effective degree of freedom was the constituent counting rule (CCR) \cite{Brod_Farrar, Brod_Farrar2}. The rule states that for hard, exclusive processes the differential cross section should have the form
\begin{equation}\label{CCR}
\frac{d \sigma}{dt} = \frac{f(\cos \theta_{c.m.})}{s^{n-2}} = \frac{f(\cos \theta_{c.m.})}{s^{N}},
\end{equation}
where $\theta_{c.m.}$ is the center-of-momentum scattering angle{\footnote{From here on, any mention of the angle $\theta$ should be taken to be in the CM frame.}}, $s$ and $t$ are Mandelstam variables, $N$ is the scaling parameter and $f$ is some function which, at least in principle, is calculable via QCD.  Here $n = \sum_i n_i$ is the sum of the total number of constituents taking part in the hard sub-process. For elementary particles $n_e=1$, for mesons $n_{M} = 2$ and for baryons $n_B = 3$. The rule is thought to be correct when the hardness of the scattering is larger than the mass squared of any of the external particles $ t  \gtrsim \max_i (M_i^2)$ and at high energy $ s  \gg \max_i (M_i^2)$ (up to small corrections, such as logarithmic corrections due to renormalization). 

\begin{figure}
\includegraphics[width=0.5\columnwidth]{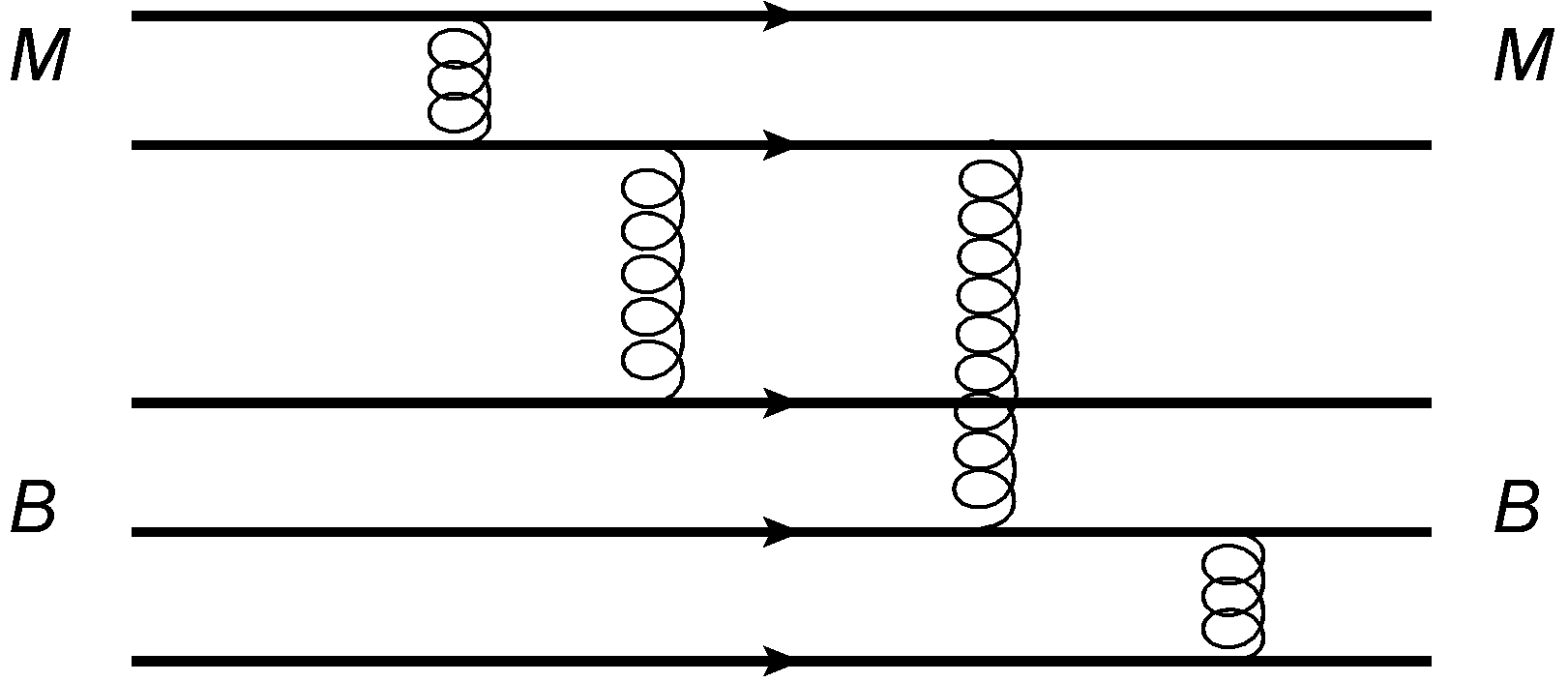}	
\centering
\caption{Example of minimal diagram connecting all constituents for $MB \rightarrow MB$. Here there are $n = 10$ constituents/external quarks, $n/2-1 = 4$ gluon propagators and $n/2-2 = 3$ quark propagators.}
\label{CCR_example}
\end{figure}

The CCR can be justified on the basis of perturbative QCD (pQCD) and dimensional arguments. Specifically, in the QCD Feynman rules for scattering amplitudes  the vertices have no dimension, gluon propagators have a dimensionality of inverse mass squared $\frac{1}{M^2}$,  quark propagators have $\frac{1}{M}$, while external quarks have $\sqrt{M}$ . Looking at a reaction with the minimal number of constituents where all participate, we need $\frac{n}{2}-1$ virtual gluons, $\frac{n}{2}-2$ virtual quarks and $n$ external quarks (see Fig. \ref{CCR_example}). It follows that the dimensionality of the scattering amplitude is $[\mathcal{M}] = \frac{1}{M^{2(\frac{n}{2}-1)}} \frac{1}{M^{\frac{n}{2}-2}} M^{n/2}=\frac{1}{M^{n-4}}$, which implies $[\frac{d\sigma}{dt}] = \frac{1}{M^{2(n-2)}}$ \cite{kawamura2013determination}. For a hard, exclusive  QCD process there is only one mass scale, $\sqrt{s}$, and thus $\frac{d\sigma}{dt} \sim \frac{1}{s^{n-2}}$, with the dimensionless constant depending on the dimensionless degree of freedom,  $\cos \theta$ (alternatively,  $\frac{-t}{s}$). Additional gluon exchange and higher Fock state contributions are suppressed by additional factors of $\sim \frac{1}{s}$. The above argument makes use of the approximate conformal symmetry of QCD, while a non-perturbative derivation has been made through the use of the AdS/CFT correspondence \cite{polchinski2002hard,brodsky2010ads}.  Including QED processes to describe meson photoproduction is straightforward.  

The CCR has been found to hold for a number of reactions, often at surprisingly low energy and hardness scales. For example, in Ref. \cite{Schumacher} it was shown that for the reaction $\gamma p \rightarrow K^{+} \Lambda$, scaling of $s^{-7}$ occurs for $t$ values as low as 1.5 GeV$^2$ and $s$ as low as 5 GeV$^2$  for $\cos\theta=0$. Scaling at relatively low energy has also been observed with the photoproduction of pions. It was seen in \cite{Gao, Zhu} for $\gamma p\rightarrow\pi^+n$ and $\gamma n\rightarrow\pi^-p$ for $s$ approximately greater than 6.25 GeV$^2$ with $\theta=90^\circ$.  The energy values (in terms of $s$) that are the focus of our work have a maximum value of 8.04 GeV$^2$ and go down to between approximately 5 and 6 GeV$^2$, depending on the angle (the cuts were made in terms of $t$). We will be looking at a kinematical range comparable to these previous studies.

In spite of the examples cited above, there has been mixed success of CCR at intermediate energies and momentum transfers. A study of 10 meson-baryon ($MB \rightarrow MB$) and baryon-baryon ($BB \rightarrow BB$) exclusive reactions at $\theta_{CM} = 90^o$ and $-t \approx 5$ GeV$^2$ found only three out ten reactions had $n-2$ within $1\sigma$ of the expected result and only one other within $2\sigma$ \cite{20reactions}. Other apparent failures of the CCR led to the development of the handbag model and the use of generalized parton distributions (GPDs) in the study of deeply virtual Compton scattering \cite{DVCS-Rady,DVCS-Ji-1,DVCS-Ji-2} and hard meson production \cite{hard-meson-production-GPD}. Arguments were made for the use of the handbag model in Compton scattering off the proton with moderate momentum transfer ($ -t \leq 10$ GeV$^2 $)\cite{Compton-scatter-theory}. A subsequent experiment at Jefferson Lab showed an extra suppression of $\frac{1}{s^2}$ for Compton scattering than the CCR predicted.\cite{Compton-scattering-experiment} The results, however, were consistent with the handbag model.

Recently, there has been a discussion in the literature about the applicability, limitations, and necessary conditions for the CCR  
\cite{guo2017constituent,brodsky2018qcd}. Ref.~\citep{guo2017constituent} claims, ``...were the constituent counting rule right, it would
provide a very powerful and straightforward tool to access the valence quark structures of the exotic hadrons. But unfortunately....for 
hadrons with hidden-flavor quarks it is problematic to apply such a naive constituent counting rule." In constrast, Brodsky 
{\it et al.}~\citep{brodsky2018qcd} state  ``constituent counting rules are completely rigorous \textit{when} they are applied properly". Given that the CCR has been suggested as a tool to study exotic hadrons, a good understanding of its correct application is needed. Here, we examine the issue by analyzing the data for a specific reaction.

A na\"{i}ve application of the CCR to $\gamma p \rightarrow  \omega  p$ would suggest $ n = n_{\gamma} + n_{p} + n_{\omega} + n_{p} = 1 + 3 + 2 + 3 = 9 $, a scaling of $N = n-2 = 7$ and thus $d \sigma / dt  \sim s^{-7}$ for fixed $\cos \theta$. It is also possible that the photon oscillates into a quark/anti-quark $q\bar{q}$  pair before interacting with the proton, meaning it would have to be treated as having two constituents. Such Vector Meson Dominance (VMD) models have been successful in explaining the greater-than-expected interaction between the photon and hadrons as arising from the hadronic component of the photon interacting strongly with the target hadron. At very large $s$ this $q\bar{q}$ Fock state of the photon would be suppressed, however at intermediate values it may play a significant role. In this case we would have $n = 10$, $N= 8$ and  $d \sigma / dt \sim s^{-8}$. However, our analysis of the data is in fact more consistent with $d \sigma / dt \sim s^{-9}$. As discussed in Sec.~\ref{sec-discussion}, we provide a theoretical argument that explains why  the correct scaling power is $N=9$.

\section{Data}\label{Data}

\begin{figure}
\includegraphics[width=\columnwidth]{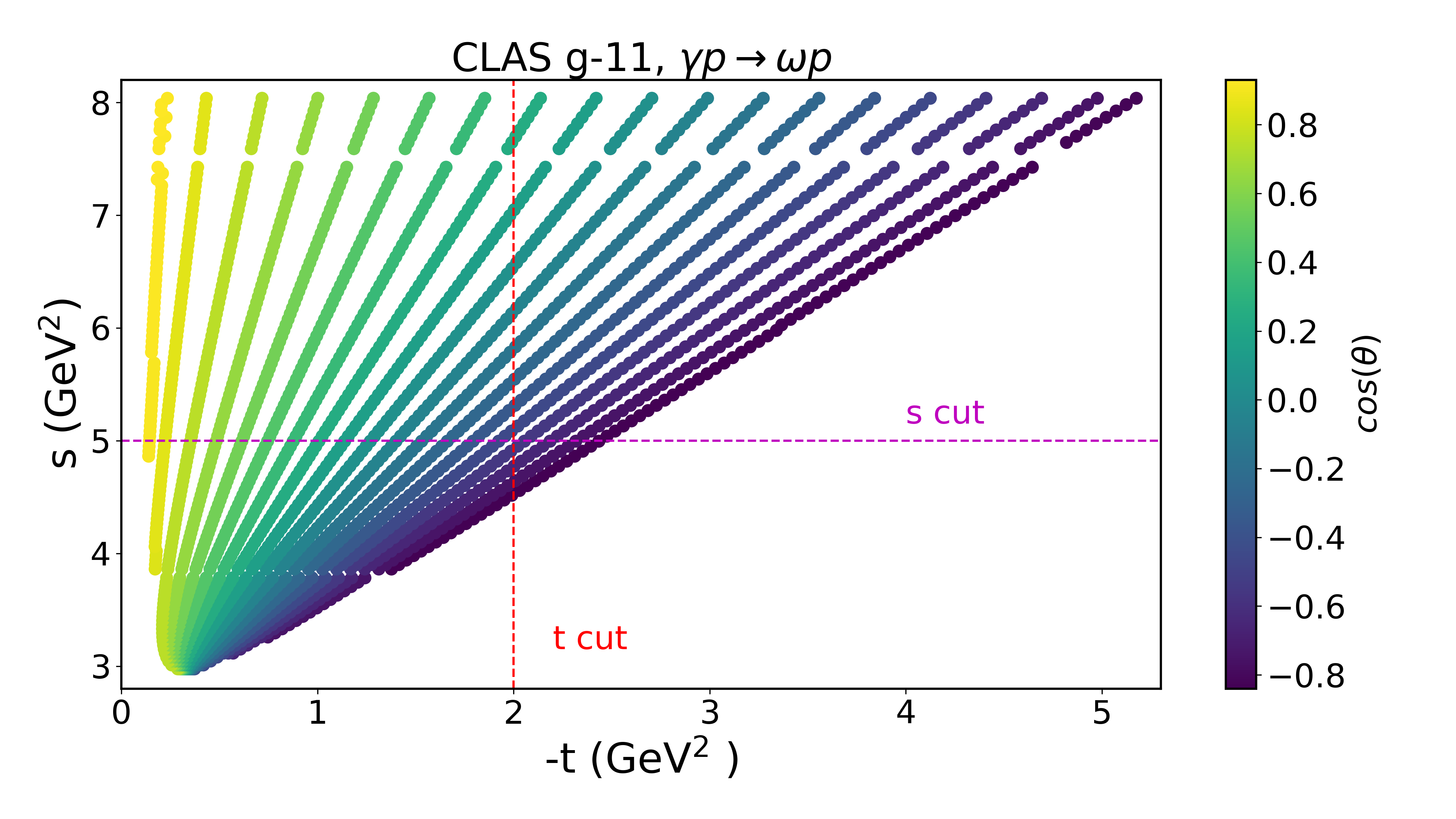}	
\centering
\caption{Kinematics of the CLAS g11 $\omega$ photoproduction data. An $s$ cut leaves low $-t$ events, but $-t$ puts a lower bound on $s$}
\label{g11data}
\end{figure}

The data being analyzed here were collected in the Jefferson Lab CLAS g11 experiment \cite{Williams}. The relatively large number of 
events and low uncertainties of the $\gamma p \rightarrow \omega p$ reaction compared to other meson photoproduction data allowed 
for an in-depth analysis. The differential cross sections in the data set were converted via:
\begin{equation}
\frac{d \sigma}{dt} = \frac{1}{2E_\gamma |\mathbf{p}_{\omega}|} \frac{d \sigma}{d\cos\theta_{c.m.}},
\end{equation}
where $E_\gamma$ is the energy of the photon and  $|\mathbf{p}_{\omega}|$ is the magnitude of the 3-momentum of the $\omega$-meson 
in the CM frame. 

The CCR is thought to hold best near $\cos\theta = 0 $ ($\theta = 90^o$) and where the $-t$ value is relatively large (hard scattering). Angles around $\theta = 90^o$ are large enough that they reduce the chances of final state interactions, while not being so large that they introduce backscattering events,  which include unwanted u-channel processes. Published experimental cross section data frequently do not have bins corresponding exactly to $\cos \theta = 0$. Such is the case with the data set used in this analysis, and thus, the four $\cos \theta$ bins $-0.15, -0.05, +0.05,$ and $+0.15$ were examined. We will show in Section \ref{Analysis} that including several angle bins can be useful in getting the scaling parameter, $N$, at $\theta = $ 90$^\circ$. Most past analyses of the CCR use a cut in Mandelstam  $s$ (i.e. $s > s_0$), in conjunction with the large center-of-momentum angle criteria to select hard scattering events.  We argue that $-t$ is the hardness of the scattering, thus should indicate the onset of pQCD.  Also, as can be seen in Fig.~\ref{g11data}, a cut in $s$ would still leave low $-t$ events, while a cut in $-t$ 
puts a lower limit on $s$. This can be seen easily in the massless limit where $-t = \frac{s}{2}(1-\cos \theta)$. A lower limit for $-t$ imposes a lower limit for $s$ since $(1-\cos \theta) \leq 2$. However, for a given $s$, $-t$ can be made arbitrarily small when $\cos \theta \rightarrow 1$. 

\begin{figure} 
\includegraphics[width=\columnwidth]{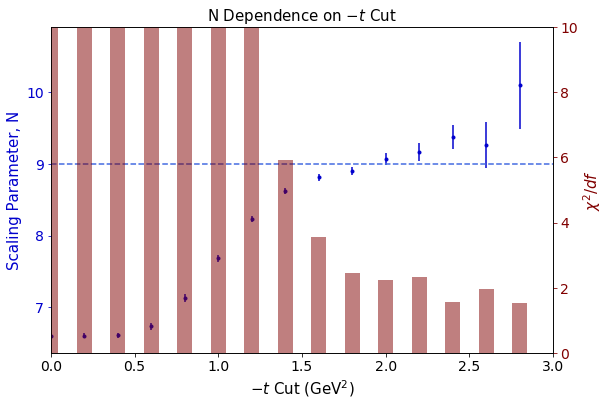}	
\centering
\caption{The value returned for the fit parameter N as a function of the $-t$ cut is shown. The fitting function used is the Taylor series expansion form of the differential cross section from Eq. (\ref{eq:3}) . The $-t$ value of each data point represents the minimum $|t|$ value allowed in the data set for that fit. The vertical bars show the reduced $\chi^2$ for each fit. The reduced $\chi^2$ values range from 83 to 13 over the $-t$ cut region of 0 to 1.4 GeV$^2$. The blue, horizontal, dashed line is added in at $N=9$ to demonstrate the scaling of this value over a range of $-t$ cuts.}
\label{N_vs_t}
\end{figure}

We selected data with $- t > 2$ GeV$^2$, resulting in 118 data points that met the criteria. The uncertainties considered were the statistical and point-to-point systematic uncertainties of the differential cross section, $\frac{d\sigma}{d\cos\theta}$, included in the published data. The $- t$ cutoff value was chosen based on a natural energy scale considering the proton mass as well as observing a plateauing of the fit parameter $N$ around $-t=2$ GeV$^2$. In this region, the exact $-t$ cut selected has minimal impact on the results. This is demonstrated by Fig. \ref{N_vs_t}. The range over $-t$ from about 1.5 to 2.5 GeV$^2$ has a relatively flat distribution of the scaling parameter, $N$. This plot also shows how poor the fits are, in terms of the $\chi^2/df$, when the $-t$ cut is less than 1.5 GeV$^2$.

\section{Analysis} \label{Analysis}
\subsection{Scaling}

\begin{figure}
\includegraphics[width=\columnwidth]{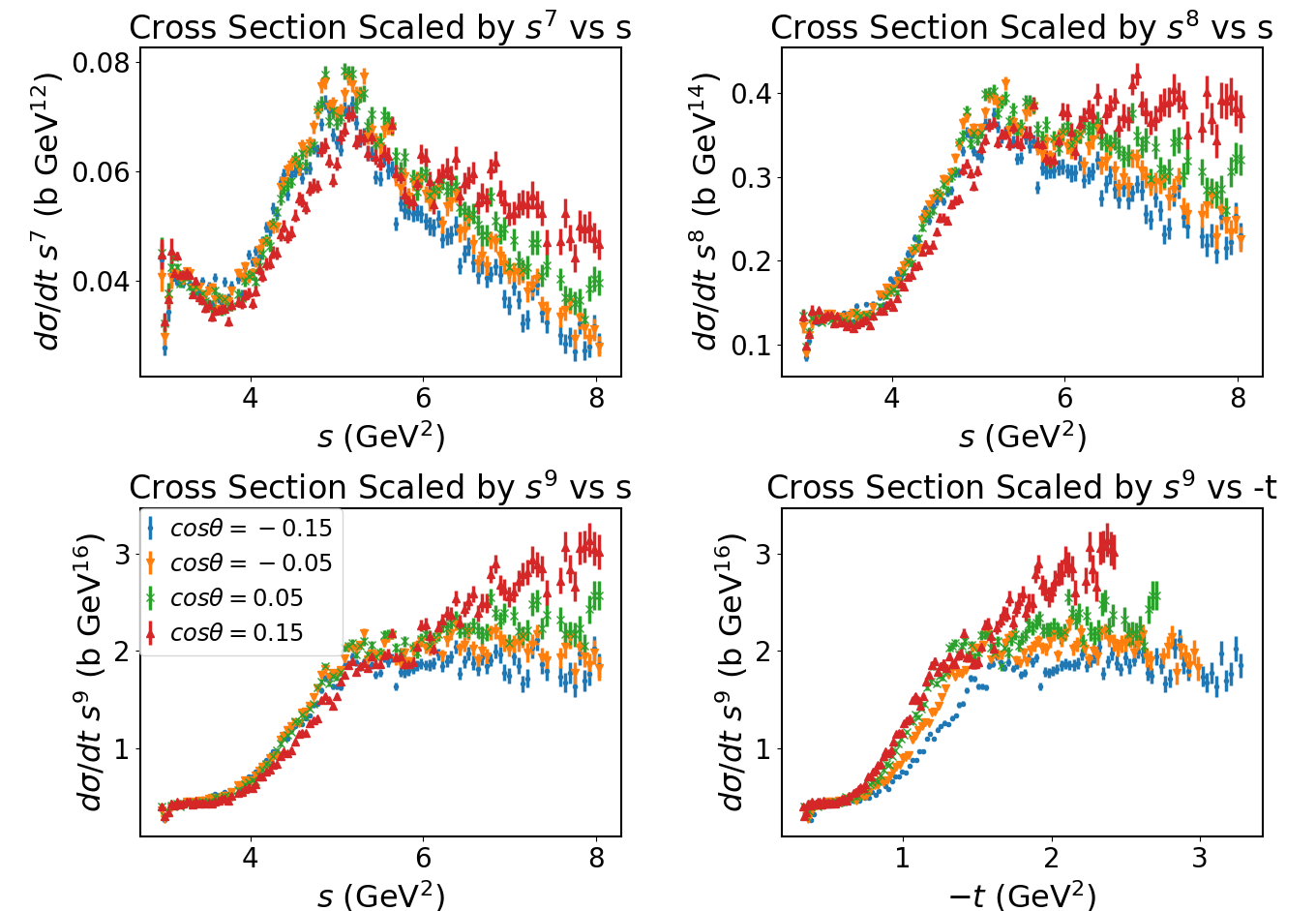}	
\centering
\caption{The differential cross sections of the four angle bins nearest $\cos\theta = 0$ with three different scaling factors applied. $s^{9}$ 
scaling is shown for both $s$ and $-t$ on the $x$-axis.} 
\label{Scaled_Cross_Sections}
\end{figure}

In Fig.~\ref{Scaled_Cross_Sections} the four angle bins around $\cos\theta=0$ are shown with various scaling factors applied. Scaling of 
$s^{-7}$ is clearly not observed for any of the angle bins. There appears to be some scaling like $s^{-8}$ above $s=6$ GeV$^2$, but only 
for the  $\cos\theta=-0.15$ bin.  The scaling 
is most obvious in the $s^{-9}$ case, where the distribution is quite flat for three of the four angles. Most importantly, the scaling is most 
extreme at the two $\cos\theta$ bins closest to 0: $\cos\theta=-0.05$ and $\cos\theta=+0.05$. This demonstrates that there is scaling as low 
as $s= 5$ GeV$^2$, which agrees with observations for other reactions. Also, in scaling by $N=9$ we see that the flattening starts at different 
$s$ for different angle bins, offering further support that cutting by $s$ is not the most appropriate approach. 

The scaling is not what we would expect it to be based on simply adding up the constituents in the reaction.  Scaling by $s^{9}$ 
(corresponding to $n = 11$) was found to work better than $n=10$, as can be seen in Fig. \ref{Scaled_Cross_Sections}. For a given angle, 
the scaled $ s^{9} d\sigma / dt$ is nearly constant over the higher$s$ ranges, while without scaling the cross section spans nearly two orders 
of magnitude. 

\subsection{Fits}
The different angle bins can be fit with one function without explicit knowledge of $f(\cos\theta)$ by noting near $\cos \theta = 0$ a Taylor 
series expansion can be taken: $f(\cos\theta) = A + B\cos\theta  + \mathcal{O}(\cos^2 \theta ) $. Keeping just the linear term we can fit the 
function, 
\begin{equation} \label{eq:3}
\frac{d \sigma}{dt} = (A + B \cos \theta) s^{-N}.
\end{equation}
Through this approach, data from multiple bins are being used. Three different types of fits were performed on the data: $\chi^2$ minimization,
Bayesian estimation, and Bootstrapping. No further details will be discussed for the $\chi^2$ minimization method, given its ubiquitousness. 
However, the following will provide a brief explanation of how the two former fitting techniques were done.

\subsubsection{Bayesian Estimation}
In the Bayesian approach one conditions on the data, $\mathcal{D}$. The posterior distribution for the parameters $A,B,$ and $N$ is
proportional to the likelihood multiplied by the prior:
\begin{equation}
\mathcal{P}(A, B ,N | \mathcal{D}) \propto \mathcal{P}(\mathcal{D}|A,B,N) \mathcal{P}(A,B,N)
\end{equation}
Since $N$ is the parameter of interest, one can then find the $N$ posterior distribution , $\mathcal{P}(N)$, by marginalizing out $A,B$. For 
the likelihood we take $\mathcal{P}(\mathcal{D}| A, B ,N) = \exp(-\frac{1}{2} \chi^2)$.

\begin{figure}
\includegraphics[width=\columnwidth]{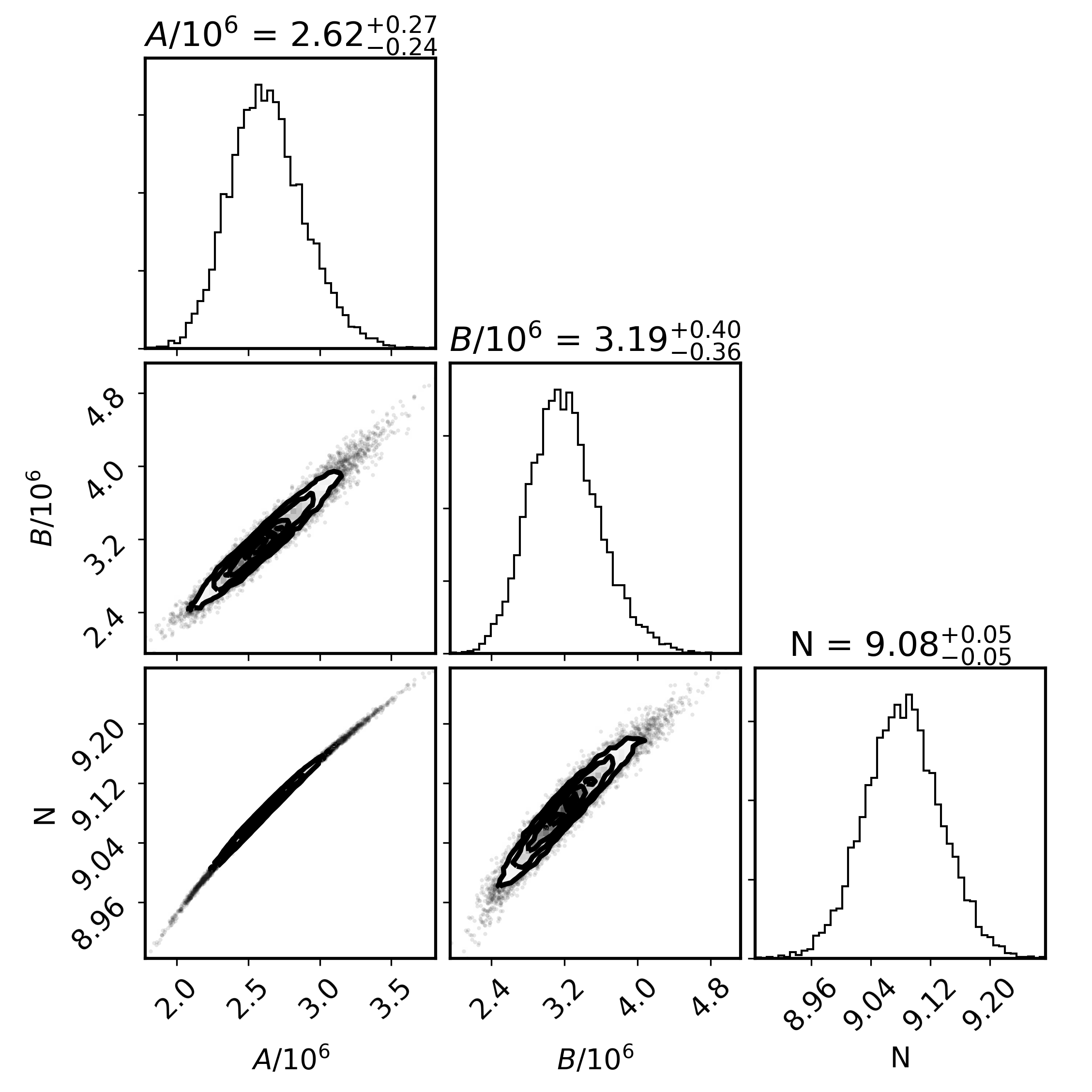}	
\centering
\caption{Results of MCMC sampling. The one-dimensional plots show the histogram of each parameter and the two-dimensional plots show the distributions in terms of each pair of parameters. Numbers above the parameter histograms indicate median and quintiles.} 
\label{Bayes_parameter_hist}
\end{figure}

\begin{figure}
\includegraphics[width=\columnwidth]{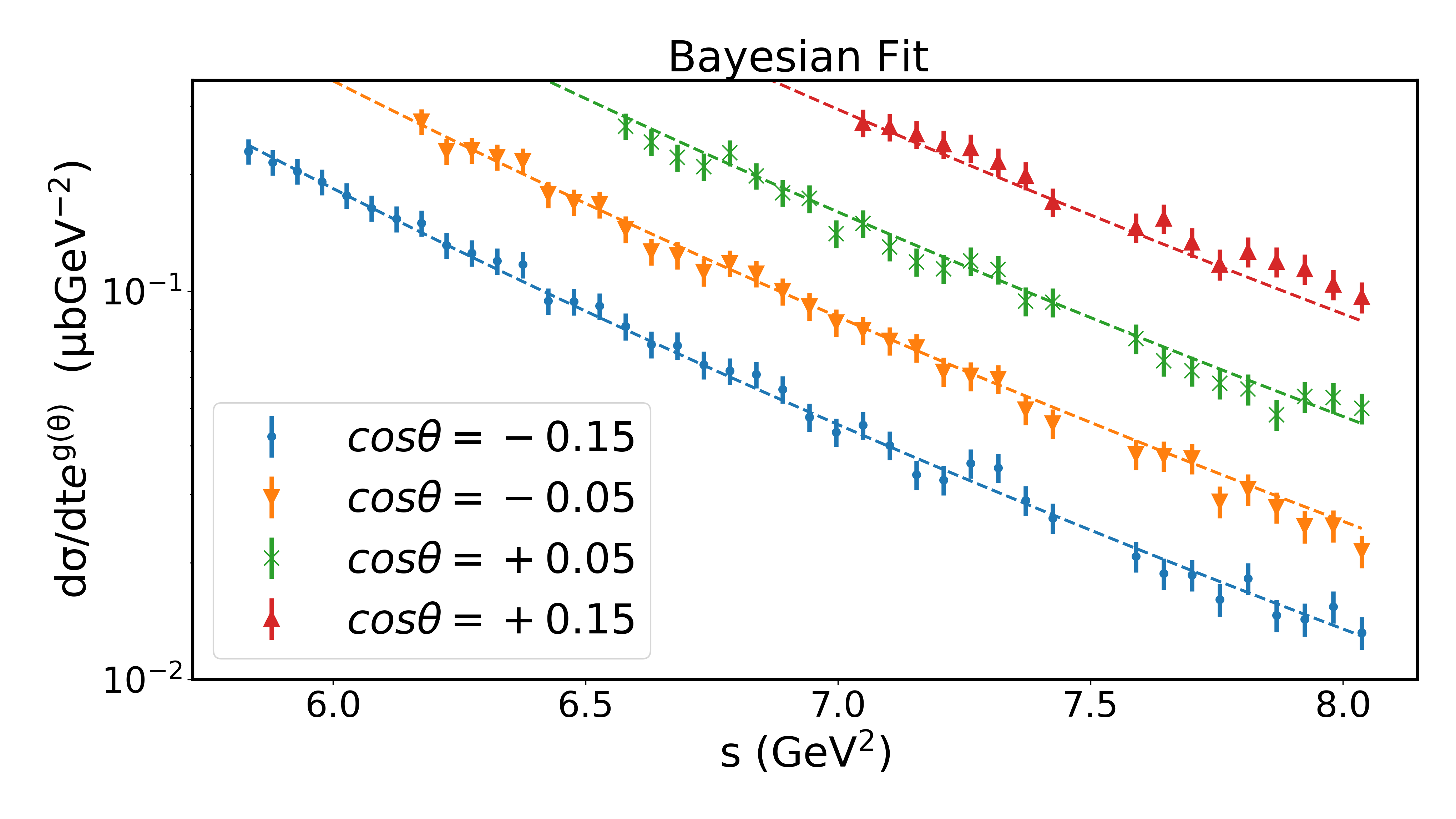}	
\centering
\caption{The differential cross section fitted with $(A + B \cos \theta) s^{-N}$, seen by the dash line. The data and fits were multiplied by $e^{g(\theta)}$  with $g(\theta) = 5(\cos\theta + 0.15)$ for readability.} 
\label{Bayesian_fit}
\end{figure}

First we assumed the 3 parameters to be independent, $\mathcal{P}(A,B,N) =\mathcal{P}(A) \mathcal{P}(B) \mathcal{P}(N)$. While not
completely realistic, it's a starting point and with enough data a more accurate relation between the variables should emerge. Uniform 
distributions were used for all three variables, where for $N$ we tried to capture our initial expectation that it should be around 7 or 8 but 
also allow some leeway by having it uniformly distributed between 5 and 10, $N \sim \mathcal{U}(5,10)$. For $A$ and $B$, we use 
$\mathcal{U}(10^4,10^7)$. The likelihood function is the same as with the $\chi^2$ approach. To get a representative sampling of parameter 
space we used a Markov Chain Monte Carlo (MCMC) algorithm implemented through the library \texttt{emcee} \cite{foreman2013emcee}. 
The lower right panel of Fig.~\ref{Bayes_parameter_hist} shows the marginalized posterior distribution for $N$ and from the mean and 
standard deviation we get the Bayesian uniform prior estimate: $N_{BU} = 9.08 \pm 0.05$. Fig.~\ref{Bayesian_fit} shows the fit using the 
average values for the parameters.

To test the dependence on the prior we also tried a Gaussian distribution for $N$ with the mean and standard deviation based on an earlier 
result \cite{Battaglieri},  $N \sim \mathcal{N}(\mu = 7.2, \sigma^2 = 0.7^2)$. We obtained nearly identical results, suggesting our results are 
not sensitive to the choice of priors.

\subsubsection{Bootstrapping}
The two methods above assume the likelihood is a Gaussian while in the Bayesian method we employed parametric distributions for the 
priors. As a check against these assumptions, we also made use of the non-parametric bootstrap method \cite{Efron1979}. Each data point was resampled
according to a uniform distribution centered at the value and a range of $\pm 2$ times the uncertainty. The result was consistent with the 
two other fitting methods. Table \ref{N_estimate} shows the results for the different methods.
\addtolength{\tabcolsep}{+10pt} 
\begin{table}
\sffamily
\setlength{\tabcolsep}{7.5pt}
\caption{Scaling Parameter Estimates}
\begin{tabular}{llll}\label{N_estimate}
Method                                 & $N$             & Comments &   \\
\toprule
$\chi^2$ Minimization                   &  $9.07 \pm 0.08$ &   $\chi^2 / df = 2.23$       &   \\
Bayesian                               & $9.08 \pm 0.05$  &   Uniform priors        &   \\
Bootstrap                              & $9.07 \pm 0.06  $        &  Uniform distribution         &  
\end{tabular}
\end{table}

\subsection{Angle and Cutoff Dependence of Scaling}
Above it was assumed that $N$ is independent of $\cos \theta$. Renormalization arguments would suggest that $N$ has some dependence on the transverse momentum transfer \cite{sivers1988nuclear}, thus on $\cos \theta$ {\footnote{A similar situation is expected in the large Bjorken $x$ region for parton distribution functions (PDFs), where it is expected that the valence PDFs go like $f(x) \sim (1-x)^N$ with $N=3$ at low $Q_0^2$ but increases with $Q^2$ \cite{holt2010distribution}.}. 

To see the effects, we broadened the inclusion of angles to $-0.3< \cos \theta < +0.3$ and for each bin took a fit of the form $As^{-N}$. 
Fig.~\ref{N_ang_dependence} shows how $N$ depends on $\cos \theta$. The $N$ value peaks near $\cos \theta = 0$, justifying our 
approximation that the scaling is independent from $\cos \theta$, although it is centered at $\cos \theta \approx -0.08$ rather than 
$\cos \theta = 0$. In the high energy limit where everything is effectively massless, the transverse momentum of the reaction is $p_\perp^2 = \frac{tu}{s} = \frac{s}{4} \left(1 -\cos^2\theta \right)$. Letting $\Lambda_{QCD}$ be the QCD scale, then from renormalization and QCD one would then expect to see quantities to evolve in terms of
\begin{equation}
\log \left(\frac{p_\perp^2}{\Lambda_{QCD}^2}\right) \simeq \log\left(\frac{s}{4\Lambda_{QCD}^2}\right) + \cos^2\theta + \mathcal{O}(\cos^4\theta),
\end{equation}
which would explain the approximate $\cos^2\theta$ dependence of $N$. Doing a quadratic fit for $N$ in terms of $\cos \theta $ suggests $N( \cos \theta = 0) \approx 9.1$, consistent with our results and offering another way to estimate the scaling parameter at $\cos \theta = 0$ using multiple angle bins. 
Taking the four bins nearest $\cos \theta = 0$ we estimate the uncertainty due to the angle dependence to be $\delta N_\theta = \frac{1}{\sqrt{\sum_i \frac{1}{\delta N_i^2}}} = 0.051$. A cutoff of $-t = 2.0$ GeV$^2$ was used. Table \ref{Unc_table} summarizes the uncertainty results.

%
\begin{figure}
\includegraphics[width=\columnwidth]{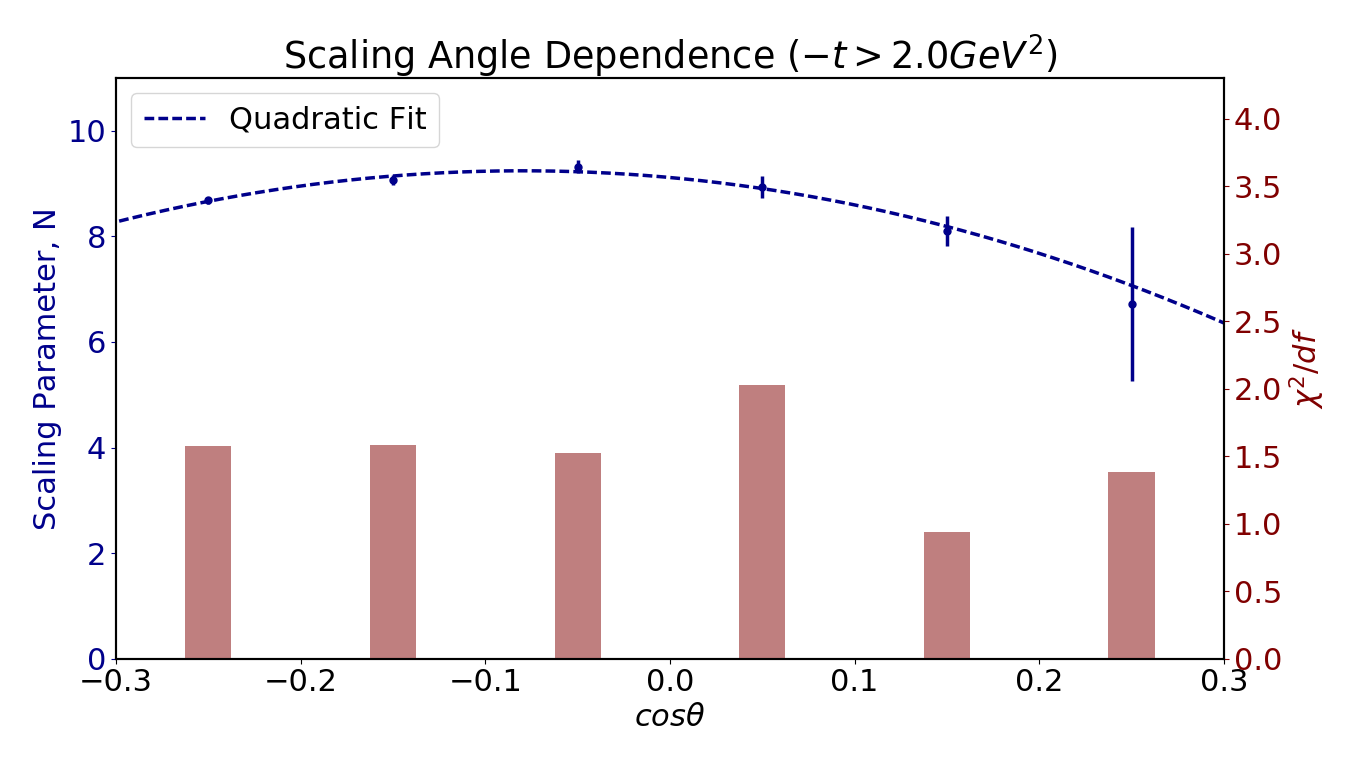}	
\centering
\caption{Dependence on $N$ with $\cos\theta$. Each bin was fitted with the function $A s^{-N}$.}
\label{N_ang_dependence}
\end{figure}

\addtolength{\tabcolsep}{+10pt} 
\begin{table}
\sffamily
\setlength{\tabcolsep}{10pt}
\caption{Fit Systematic Uncertainties}
\begin{tabular}{llll}\label{Unc_table}
Source of Uncertainty                  & $\delta N$ Estimate           & \\
\toprule
Fit                                    & $0.051$ &    \\
Angle Dependence                       & $0.067$  &      \\
$-t$ Cutoff Dependence                      & $0.072 $        & 
\end{tabular}
\end{table}

\section{Discussion} \label{Discussion}
\label{sec-discussion}

The results obtained here contradict an earlier study that found $N=7.2\pm0.7$ \cite{Battaglieri}. However, that result was based on only 
5 data points and all the points (except for one that was used from SLAC that has  $s= 6.13$ GeV$^2$) have $s>7$ GeV$^2$,
going up to $s= 8.06$ GeV$^2$. Although these values are within the same range we are examining, they are not representative of the 
overall data set used in this analysis. The data discussed in this paper that pass the $-t > 2 $GeV$^2$ cut have $s$ values that 
go down to as low as approximately 5.8 GeV$^2$, depending on the angle. It is, however, fairly consistent with another analysis that 
found a scaling of $N = 9.4 \pm 0.1$ \cite{dey2014scaling}. As mentioned before, a Compton scattering experiment at Jefferson Lab also found a extra factor of $\frac{1}{s^2}$ from what was expected by CCR \cite{Compton-scattering-experiment}. 

One possible explanation for the discrepancy between the na\"ive application of the CCR and the results of this analysis} is that $s$ and $-t$ are too low for it to be applicable. However, as 
mentioned earlier, the CCR does seem to explain other reactions at comparable energies and hardness scales. Also, this argument does 
not explain why it \textit{does} work with $N = 9 $. 

\begin{figure}
 	\centering
 	\includegraphics[width=0.8\linewidth]{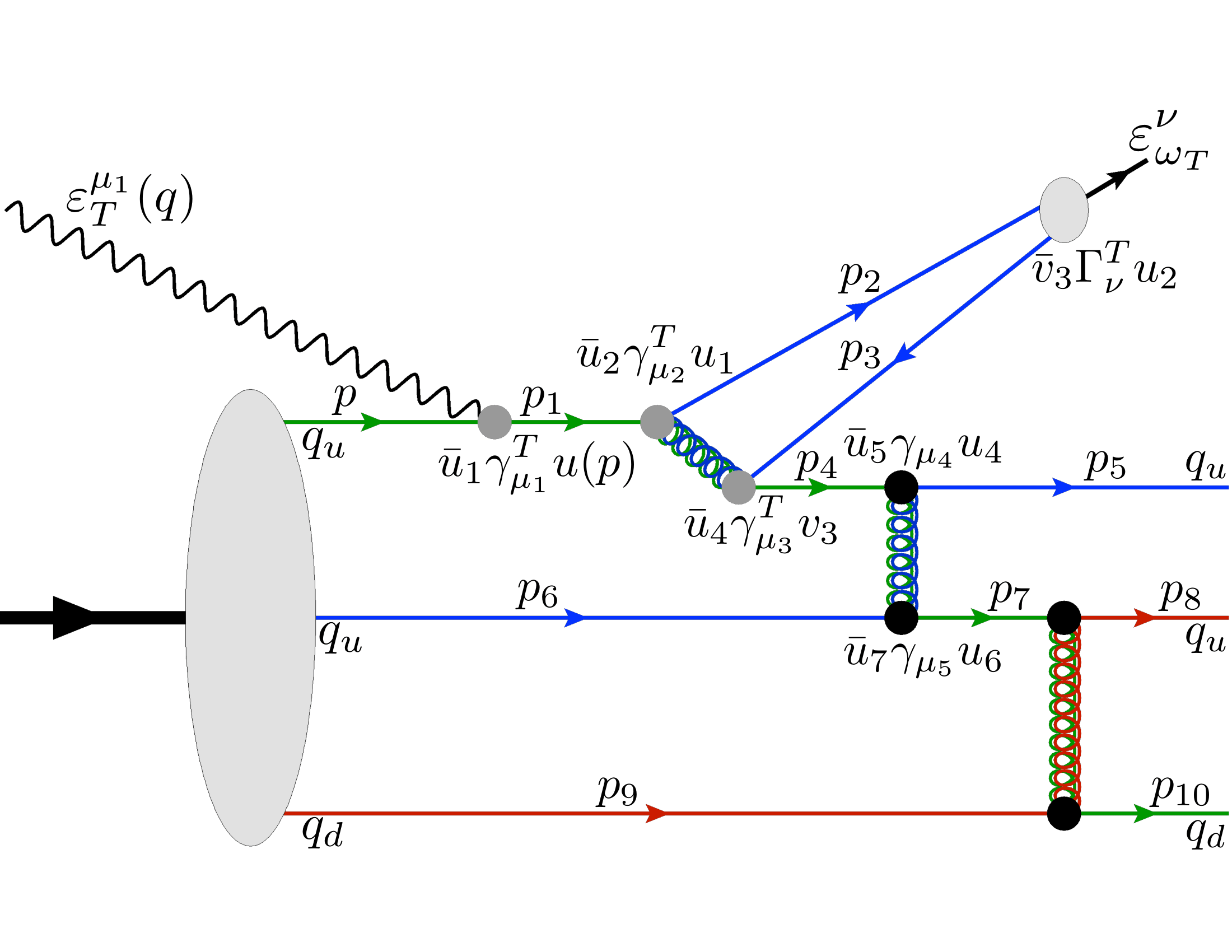}
 	\caption{Diagram for $\omega$ photoproduction.  
 	The polarization vectors of the photon	$\varepsilon_T^{\mu_1}$ and vector meson $\varepsilon_{\omega_T}^{\nu}$ are constrained 
	to be  transverse. }
 	\label{omegahardscatt}
 \end{figure}

We propose the following mechanism to explain the results.
In Fig.~\ref{omegahardscatt} a model for $\omega$ photoproduction off the proton 
is presented. 
Since the reaction is at a high enough energy, one can assume helicity conservation of quarks participating in the hard scattering sub-process. The matrix 
elements with vertices  $\gamma^T_{\mu_i}  \ (i=1,2,3 )$, are constrained to be transverse because they coupled to vector bosons (photon 
and meson) with transverse polarization states.

Consider the photon vertex $\gamma^T_{\mu_1}$, which in the Breit frame of $u(p)$ there is no transverse momentum ($p_\perp=0$), thus 
no orbital angular momentum to account for. Since, the momentum flips, $\vec{p}=-\vec{p_1}$,  helicity conservation implies that the spin 
must also flip. 
Using Tables II and III from Appendix A of Ref.~\cite{PhysRevD.22.2157},  we can see  there is a suppression of one power of $s$ at each of 
the vertices $\gamma_{\mu_1}$ and $\gamma_{\mu_2}$. For $\gamma_{\mu_3}^T$ there is no suppression if we demand 
$p^T_{2} \sim p^T_{3} \sim p^+_{2} \sim p^+_{3} $, i.e. $t \sim s $, which happens at large angles. If $t \sim s$ is not fulfilled there is an 
extra power suppression at $\gamma_{\mu_3}^T$ as well. Overall, the power counting is of two powers of $s$ more than the naive one, 
leading to, $N=11-2=9$. Thus, a careful consideration of conservation of angular momentum to the CCR explains why $d\sigma/dt$ in the photoproduction of the spin-1 $\omega$ behaves differently from that of scalar mesons. Given that the $\rho^0$ meson is also spin-1 and has a similar quark structure to the $\omega$, we predict a similar scaling should follow.

\section{Conclusion}\label{Conclusion}
We have examined the published CLAS g11 data for $s$ scaling in $\omega$ photoproduction. Our analysis is unique in cutting by the 
relevant hardness parameter $-t$ (rather than $s$) and combining several angle bins through a Taylor expansion of the 
$\cos \theta$ dependent function of $\frac{d\sigma}{dt}$.  Around $\theta = 90^\circ$, we found the scaling to be $N = 9.08 \pm 0.11$,  inconsistent with a na\"{i}ve application of the CCR.  We have made the case for additional gluon exchanges and spin-flipping to conserve helicity. Additional exploration of these mechanisms are warranted. Future CLAS12, GlueX, and the CLAS g12 experiment ($E_\gamma$ up to 5.45 GeV) data for this reaction at higher energy scales could be further used to examine if this trend continues. Other reactions from these experiments could also be tested against our observations. Our results suggest that the CCR is in fact valid for the examined data, however the data selection must be based on hardness of the scattering using $-t$ instead of $s$. 
It is thus worth pointing out that when applied appropriately, the CCR could also be used to investigate exotic hadrons such as hybrid mesons 
to be studied in the GlueX experiment~\cite{GlueX2010}, where a range of beam energies will make an analysis similar to ours possible.

\section*{Acknowledgments}
This research was funded in part by the U.S. Department of Energy, Office of Nuclear Physics under contracts No. DE-SC0013620 and 
DE-FG02-01ER41172. We would like to thank Werner Boeglin and Misak Sargsian for useful discussions.  


\end{document}